\documentstyle[aps,epsf,floats]{revtex}  

\newcommand{\etal} {et~al.}

\begin{document}        

\baselineskip 14pt
\title{A Search for CP(T) Violation in B Decays at OPAL}
\author{Robin Coxe}
\address{Enrico Fermi Institute, University of Chicago}

%
\maketitle              

\begin{abstract}        
A search for CP(T) violation is performed and the fractional
difference between the ${\rm b}$ and ${\rm \bar b}$ hadron lifetimes is 
measured using reconstructed secondary vertices in inclusive B hadron
decays selected from 3.1 million ${\rm Z^0 \rightarrow q \bar q}$ events.
The data were collected by the OPAL experiment at the LEP collider at CERN
at ${\rm \sqrt{s} \approx 91}$ GeV from 1991-1995.  
\
\end{abstract}   	

\section{Introduction}               

CP violation in the B-meson system has
generated considerable experimental and theoretical interest, as
potentially large effects are expected.  Searches for CP(T) violation
using the small sample of ${\rm Z^0 \rightarrow b \bar b}$ decays
at the LEP collider at CERN provide ``proofs of principle" for 
analysis techniques which will be employed by future B-Factory
experiments. 

Indirect CP violation is possible in the ${\rm B^0}$ system provided the
weak eigenstates ${\rm B^0}$ and ${\rm \bar B^0}$ differ from the mass
eigenstates ${\rm B_1}$ and ${\rm B_2}$:
\begin{eqnarray}
{\rm |B_1>} & = & {{
(1+{\epsilon_B}+{\delta_B}){\rm |B^0 >} +
(1-{\epsilon_B}-{\delta_B}){\rm |\bar B^0 >} } \over
{\sqrt{2 \left(1 + \left| {\epsilon_B} +{\delta_B}
\right|^2
\right) } } }
\\
{\rm  |B_2>} & = & {{
(1+{\epsilon_B}-{\delta_B}){\rm |B^0 >} -
(1-{\epsilon_B}+{\delta_B}){\rm |\bar B^0 >} } \over
{\sqrt{2 \left(1 + \left| {\epsilon_B} -{\delta_B}
\right|^2
\right) } } },
\end{eqnarray}
where $\epsilon_B$ and $\delta_B$ parametrize CP and CPT violation,
respectively\cite{kp}.  These parameters have been investigated
using semileptonic b hadron decays, resulting in limits of order $10^{-2}$
on both $\epsilon_B$ and $\delta_B$\cite{rvk,oldeb,opaldms,rjh}. In the
Standard Model,  Re(${\rm \epsilon_B}$) is expected to be
around $10^{-3}$\cite{cpinc}, but it could be up to an order of
magnitude larger in superweak models\cite{superw}.

A non-zero value of ${\rm \epsilon_B}$ gives rise to a time-dependent rate
asymmetry, ${\rm A(t)}$,  in inclusive ${\rm B^0}$  {\it vs.} inclusive
${\rm \bar B^0}$ decays\cite{cpinc}, defined as:
\begin{eqnarray}
{\rm A(t)} & \equiv & {{ {\rm B^0(t)}  - {\rm \bar B^0(t)}}
\over { { \rm B^0(t)}  + {\rm \bar B^0(t)} } },
\label{eq:asym1}
\end{eqnarray}
where ${\rm B^0(t)}$ and ${\rm \bar B^0(t)}$ are the
decay rates of ${\rm B^0}$ and ${\rm \bar B^0}$ mesons.  
For an unbiased selection of ${\rm B^0}$ and ${\rm
\bar B^0}$ mesons, the time-dependent inclusive decay rate asymmetry can
be rewritten in terms of proper decay time ${\rm t}$:
\begin{eqnarray}
A(t)  & = & a_{cp} \left[ {{\Delta m_d \tau_{B^0} } \over
{2}}
\sin (\Delta m_d t ) - \sin^2 \left( {{\Delta m_d  t}
\over {2}} \right) \right], 
\label{eq:asym2}
\end{eqnarray}
where ${\rm a_{cp}}$ is the CP-violating observable, ${\rm
\Delta m_d}$ is the ${\rm B^0}$ oscillation frequency, and
${\rm \tau_{B^0}}$ is the ${\rm B^0}$ lifetime. \\ For 
${|\epsilon_B|<<1}$, the parameter ${\rm a_{cp}}$ is related to
${\epsilon_B}$ by: ${\rm Re(\epsilon_B) =  a_{cp}/4}$.

Furthermore, CPT invariance implies that ${\tau_{b} = \tau_{\bar b}}$.
If CPT violation occurred, the lifetimes of ${\rm b}$ and ${\rm \bar b}$ 
hadrons could be different:
\begin{eqnarray}
\tau_{b/ \bar b} & = & \left[ 1 \pm {{1}\over{2}} \left(
{{\Delta \tau}\over{\tau}} \right)_b \right] \tau_{av},
\end{eqnarray}
where ${\rm \tau_{av}}$ is the average and ${\rm (\Delta
\tau/\tau)_b}$ is the fractional difference in lifetimes.

\newpage
\section{Inclusive CP(T) Tests}
The measurement of the time-dependent rate asymmetry, A(t), and the 
extraction of Re($\epsilon_B$) proceeds in several steps.  First,
selected ${\rm Z^0 \rightarrow q \bar q}$ events are divided into 2
hemispheres defined by the plane $\bot$ to thrust axis and containing the
${\rm e^+ e^-}$ interaction point.  A sample of about 400,000 ${\rm Z^0
\rightarrow b \bar b}$ events is identified using b-tagging techniques
described in detail in~\cite{opalrb,elecid,muonid,opaldil}.  The b-tags
rely on the presence of a displaced secondary vertex or a high momentum
lepton.  In each event, the hemisphere containing the b-tag is referred to
as the ``T-tagged" hemisphere. The b-tag has a ${\rm \epsilon_{hemi}}$ of
37 \% for ${\rm b \bar b}$ events and a  non-b impurity of 13\%.  Next,
the b hadron proper decay time, t, in the opposite ``measurement"
hemisphere is reconstructed by forming a secondary vertex, measuring the
decay distance from the primary vertex, and estimating the b hadron
energy. The quantity ${\rm a_{cp}}$ is then extracted via a binned
$\chi^2$-fit to the observed time-dependent asymmetry in bins of
reconstructed proper time.

\subsection{Production Flavor Tag}

The production flavor estimate, ${\rm Q_T}$, in the tagged hemisphere is
the output of a neural net with the following inputs:
\begin{itemize}
\begin{enumerate}
\item Jet charge of the highest energy jet, ${\rm Q_{jet}}$, with (with
momentum weight $\kappa=0.5$). The jet charge is defined as:
\begin{eqnarray}
Q_{jet} & = & { {\Sigma_i (p_i^l)^\kappa q^i} \over {\Sigma_i
(p_i^l)^\kappa q^i}}, 
\end{eqnarray}
where $p_i^l$ is the longitudinal momentum component with respect to the 
jet axis and $q_i$ is the charge of track $i$.  
\item Vertex charge, $Q_{vtx} = \Sigma_i w_i q_i$, where $w_i$ is
the weight for track $i$ to have come from a secondary instead of a
primary vertex.  Weights are derived from an artificial neural network
with three inputs: the momentum of track $i$, the
transverse momentum of track $i$ with
respect to the jet axis, and the impact parameter. 
\item Uncertainty on the vertex charge, $\sigma_{Q_{vtx}}=\Sigma_i 
w_i ( 1- w_i) q_i^2$.
\item The product of lepton charge and output of neural
network used to select ${\rm b\rightarrow l}$ decays, a {lepton tag}. 
\end{enumerate}
\end{itemize}

The jet charge and vertex charge are not charge symmetric due to
detector effects resulting in the difference in the rate and
reconstruction of positively and negatively charged tracks.  Offsets are
evaluated using inclusive samples of T-tagged events and are
subtracted prior to constructing ${\rm Q_T}$.  The lepton tag
is diluted by ${\rm b}$ mixing, cascade ${\rm b}$ decays, ${\rm c}$
decays, and fake leptons. Separate networks with fewer inputs are trained
for events with a vertex or a lepton only.  

The variable ${\rm Q_T}$ is defined as:
\begin{eqnarray}
Q_T & = & {{N_b(x) - N_{\bar b}(x) } \over  {N_b(x) + N_{\bar b}(x) }} \\
|Q_T| & = & 1- 2 \eta, 
\end{eqnarray}
where ${\rm N_b(x)}$ and ${\rm N_{\bar b}(x)}$ are the numbers of MC
${\rm b}$ and ${\rm\bar b}$ hadron hemispheres with a
given value of artificial neural network output $x$ and $\eta$ is the
mistag probability.  If  ${\rm Q_T > 0}$, then the tagged
hemisphere is more likely to contain a ${\rm b~hadron}$ than a 
${\rm \bar b~hadron}$, and vice versa.  If ${\rm Q_T = 0}$, both
hypotheses are equally likely.

The ${\rm Q_T}$ flavor tag has some sensitivity to the {\it decay} flavor
of the tagged hemisphere, which is not desirable in an inclusive
measurement.  Therefore, another tag ${\rm Q_M}$ is applied in the
opposite, or measurement, hemisphere.  The ${\rm Q_T}$ output in the
T-tagged hemisphere, as well as the jet charge in the measurement
hemisphere, ${\rm Q_M}$ (with momentum weight $\kappa=0$), are combined to
construct the composite variable:  
\begin{eqnarray} 
Q_2 & = & 2 \left[{{(1-Q_T)(1+Q_M) } \over 
{ (1-Q_T)(1+Q_M) + (1+Q_T)(1-Q_M) }} \right] -1. 
\end{eqnarray} 

Again, if ${\rm Q_2>0}$ (${\rm Q_2<0}$), the so-called ``M-tagged"
hemisphere contains a ${\rm b}$-hadron tag (${\rm \bar b}$-hadron) tag.  
The ${\rm Q_2}$ variable is designed to be sensitive to the production,
but not the decay flavor of the ${\rm b}$-hadron, thus avoiding biases to
the reconstructed proper time measurement.  After flavor tagging, 394119
events remain in the data sample.  

\subsection{Proper Decay Time Reconstruction}
The CP-violating parameter ${\rm a_{cp}}$ can be extracted from the rate
asymmetry distribution, A(t), as defined in 
Equation~\ref{eq:asym2}.  This is accomplished by  
calculating the number of ${\rm b}$-hadron M-tags minus the number of
${\rm\bar b}$-hadron M-tags in
bins of reconstructed proper time $t$ and performing a binned $\chi^2$
fit to measure ${\rm a_{cp}}$.  The ${\rm b}$-hadron proper time is
defined as:
\begin{eqnarray}
t & = & {{m_b L} \over { \sqrt{E_b^2-m_b^2} } },
\label{eq:propt}
\end{eqnarray}
where $L$ is the hadron decay length, $E_b$ is the ${\rm b}$-hadron
energy, and $m_b$ is the mass of the ${\rm b}$-hadron, taken be that
of the ${\rm B^+}$ and ${\rm B^0}$ (5.279 GeV)\cite{pdg98}.

The hadron decay length, $L$, is reconstructed in the measurement
hemisphere by first forming a ``seed" secondary vertex using the two
tracks with the largest impact parameter, $d_0$, relative to the primary
vertex in the highest energy jet.  All tracks with $p>0.5$ GeV, $|d_0|<1$
cm, and $\sigma_{d_0}<0.1$ cm, which are consistent with the ``seed"
vertex are then added to it via an interative procedure.  The secondary
vertex must contain at least 3 tracks and have an invariant mass exceeding
0.8 GeV, assuming all constituent tracks are pions.  To further eliminate
badly reconstructed or fake secondary vertices, the secondary vertex must
be kinematically consistent with a long-lived particle originating from
the
primary vertex.  Secondary vertices meeting the above criteria are
identified in approximately 70$\%$ of M-tagged hemispheres for both
signal and background.  The decay length $L$ between the primary and
secondary vertices is then calculated using the jet axis as a constraint.

The ${\rm b}$-hadron energy is computed by first estimating the energy
of the ${\rm b}$-jet by treating the event as a 2-body decay of a ${\rm
Z^0}$ into a ${\rm b}$-jet of mass ${\rm m_b}$ and another object. The
charged
and neutral fragmentation energy, ${\rm E_{bfrag}}$, was estimated using
the procedure described in~\cite{opalbinc}, involving the charged track
weights ${\rm w_i}$, and the unassociated electromagnetic calorimeter
clusters weighted according to their angle with respect to the jet
axis. The ${\rm b}$-hadron energy is then, ${\rm E_b=E_{bjet}-E_{bfrag}}$.

The reconstructed proper time distribution described by
Equation~\ref{eq:propt} is convolved with 2 Gaussians to account for
detector resolution effects.  The RMS widths of the resolution functions 
are 0.33 and 1.3 ps and are determined from Monte Carlo studies.  About
65$\%$ of events lie within the narrower Gaussian.  These resolution
functions represent an average over all true decay proper times ${\rm t}$.  
The non-Gaussian effects apparent in small slices of ${\rm t}$ due to
contamination from primary vertex tracks are not critical in this
analysis, as the result is not particularly dependent on accurate decay
time resolution.

\section{Fit to Re(${\rm \epsilon_B}$)}

For each of 34 time bins ${\rm i}$ (in the range -2 to 15 ps), the
asymmetry is calculated in 10 bins ${\rm j}$ of ${\rm |Q_2|}$:
\begin{eqnarray}
A_{ij}^{obs} & = & { {N_{ij}^{b} - N_{ij}^{\bar b}} \over
{ <\left| Q_2 \right|>_{ij} (N_{ij}^{b} + N_{ij}^{\bar b})} },
\end{eqnarray}
and the error $\sigma_{A^{\rm obs}_{ij}}$ is given by:
\[
\sigma_{A^{\rm obs}_{ij}}=
\frac{1-(<\left|Q_2\right|>_{ij}\,A^{\rm obs}_{ij})^2}
{2 <\left|Q_2\right|>_{ij}}\
\sqrt{\frac{N^{\rm b}_{ij}+N^{\rm\bar{b}}_{ij}}
{N^{\rm b}_{ij}N^{\rm\bar{b}}_{ij}}},
\]
where $N^{\rm b}_{ij}$ ($N^{\rm\bar{b}}_{ij}$) is the number of
events with ${\rm Q_2>0}$ (${\rm Q_2<0}$). The factor
${\rm 1/<\left|Q_2\right|>_{ij}}$ corrects for the tagging dilution
(mis-tagging), which reduces the observed asymmetry for imperfectly tagged
events.

\begin{figure}[ht]	
\centerline{\epsfxsize 5.0 truein \epsfbox{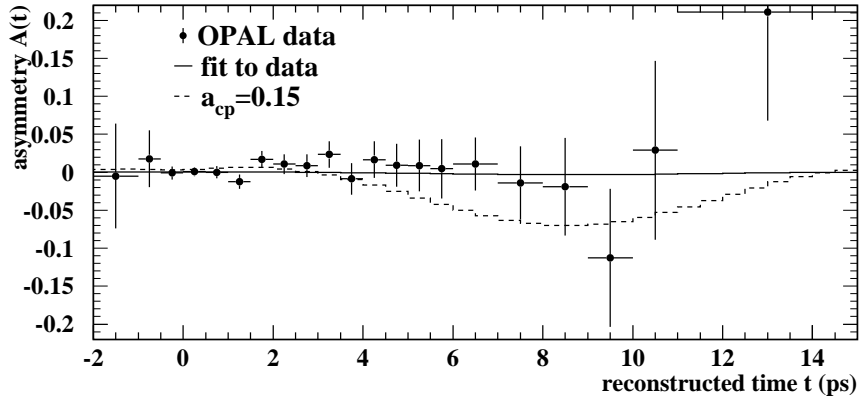}}   
\vskip .2 cm
\caption[]{Asymmetry of tagged ${\rm b}$ and ${\rm \bar b}$-
  hadrons as a function of reconstructed time $t$ in data
  (points), and fit (solid line). The expected asymmetry for
  ${\rm a_{cp}=0.15}$ is shown by the dotted line.
\label{fig:asym_fit}
}
\end{figure}

The 10 estimates of ${\rm A_{ij}^{obs}}$ in each bin
$i$ are then averaged, weighting by ${\rm (\sigma_{A_{ij}^{obs}})^{-2}}$
to get ${\rm A_{i}^{obs}}$.  A binned $\chi^2$-fit to the
reconstructed proper time which accounts for non-${\rm B^0}$ background,
the ${\rm B^0}$ lifetime, and the Gaussian time resolution functions
yields:
\begin{eqnarray}
a_{cp} & = & 0.005 \pm 0.055 \pm 0.013 \\
Re(\epsilon_B) & = & 0.001 \pm 0.014 \pm 0.003.
\end{eqnarray}
The asymmetry, A(t), as a function for reconstructed proper time ${\rm t}$
is shown in Figure~\ref{fig:asym_fit}.  The dots denote OPAL data, the
solid
line the fit result, and the dashed line the expected asymmetry for ${\rm
a_{cp}=0.15}$.

The systematic uncertainties are summarized in
Table~\ref{tab:syst}.  Detailed descriptions of the various contributions
can be found in Reference~\cite{cptpub}.  If the reconstruction efficiency
for ${\rm B^0}$ decays to different numbers of charm hadrons is not the
same, the expected asymmetry could take the form:
\begin{eqnarray}
A(t) & = & c_{cp} \sin \left( \Delta m_d t \right) - a_{cp} \sin^2 \left(
\Delta
m_d t/2 \right). 
\end{eqnarray}
Repeating the fit, letting both ${\rm a_{cp}}$ and ${\rm c_{cp}}$
vary gives:
\begin{eqnarray}
a_{cp} & = & 0.002 \pm 0.055 \\
c_{cp} & = & 0.026 \pm 0.027. 
\end{eqnarray}
Differences in efficiency are not significant, as ${\rm a_{cp}}$
does not change much. The systematic uncertainties on the measurement of
${\rm c_{cp}}$ are listed in the second column of Table~\ref{tab:syst}.

\begin{table}
\caption{Systematic uncertainties on the measurements of
${\rm a_{cp}}$ and ${\rm c_{cp}}$.}
\begin{tabular}{l|cc} 
Source & ${\rm \Delta a_{cp}}$ & ${\rm \Delta c_{cp}}$ \\
\hline
${\rm B^0}$ lifetime & 0.002 & 0.000 \\
${\Delta m_d}$ value & 0.001 & 0.001 \\
${\rm B^0}$ fraction & 0.002 & 0.002 \\
Flavor tagging offsets & 0.003 & 0.013 \\
Flavor tagging mis-tag  & 0.009 & 0.005 \\
b fragmentation & 0.008 & 0.006 \\
Time resolution & 0.002 & 0.000 \\
\hline
Total & 0.013 & 0.015 \\
\end{tabular}
\label{tab:syst}
\end{table}

\section{Fit to ${\rm (\Delta \tau/\tau)_b}$}

\begin{figure}[ht]	
\centerline{\epsfxsize 5.0 truein \epsfbox{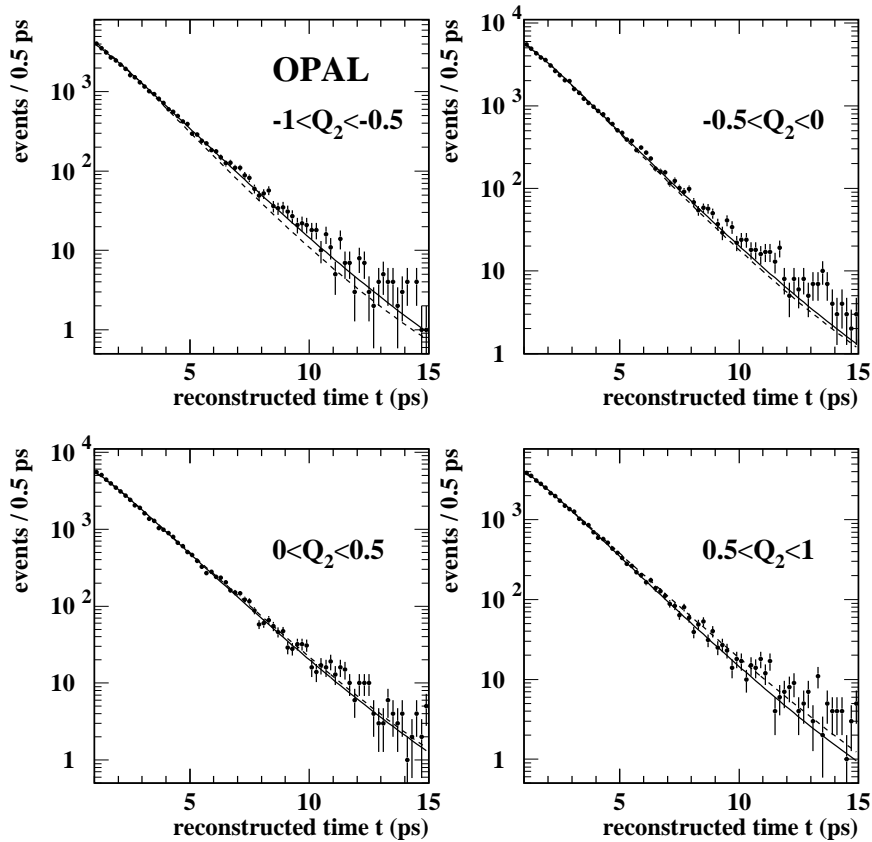}}   
\vskip -.2 cm
\caption[]{Reconstructed proper time distributions in four ranges of ${\rm
Q_2}$.  The data are shown by the points with error bars and the fit
prediction by the solid lines.  The expected distributions for  ${\rm 
(\Delta \tau / \tau)_b =0.2}$  are shown by the dotted lines.
\label{fig:rpt}
}
\end{figure}

The fractional difference between the ${\rm b}$ and ${\rm \bar b}$-
hadron lifetimes is measured by dividing the data into 20 bins of ${\rm
Q_2}$ (related to ${\rm b}$/${\rm \bar b}$-hadron purity) and performing a
simultaneous fit of the reconstructed proper time distrbutions to the
expected proper time distribution.  The fit yields values for both ${\rm
\tau_{avg}}$ and ${\rm (\Delta \tau/\tau)_b}$.  However, the result for 
${\rm \tau_{avg}}$ has a large systematic uncertainty
due to the time resolution function and should not be interpreted as a
measurement of the average ${\rm b}$-hadron lifetime. The expected
distribution accounts for time resolution effects, non-${\rm b \bar b}$
background, the lifetimes of ${\rm b}$ and ${\rm \bar b}$-hadrons, and a
background component with a lifetime ${\rm \tau_{bg}}$.
The fit result is:
\begin{eqnarray}
\left( \Delta \tau / \tau \right)_b & = &  -0.001
\pm 0.012 \pm 0.008. 
\end{eqnarray}
The uncertainty in the flavor mistag rate dominates the systematic
uncertainty on ${\rm (\Delta \tau / \tau)_b }$.  Reconstructed proper time
distributions in 4 ranges of ${\rm Q_2}$ are shown in
Figure~\ref{fig:rpt}.

\section{Conclusion}

\begin{figure}[ht]	
\centerline{\epsfxsize 5.0 truein \epsfbox{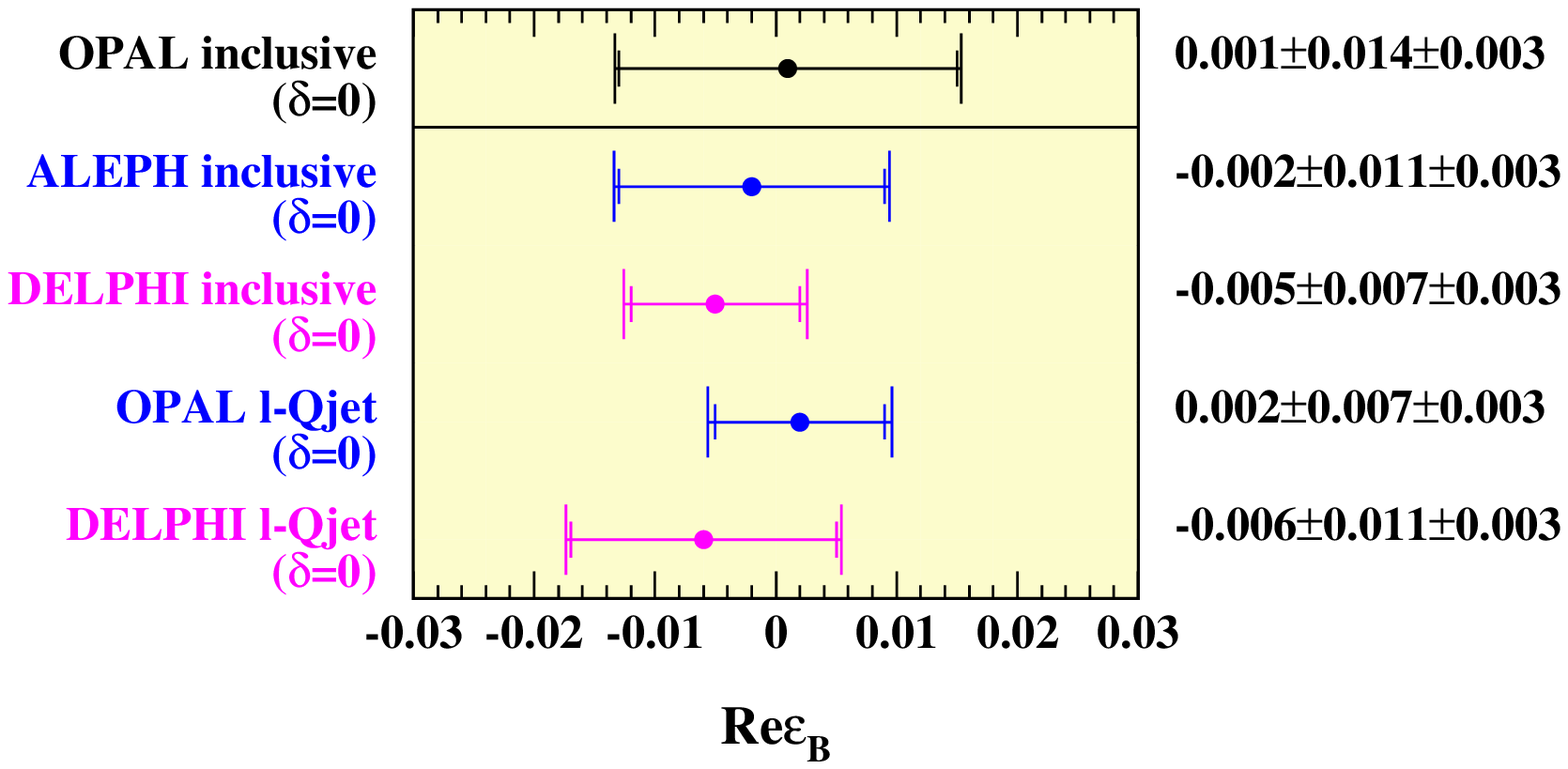}}   
\vskip .2 cm
\caption[]{
\label{fig:worldcp1}
}
\end{figure}

\begin{figure}[ht]	
\centerline{\epsfxsize 5.0 truein \epsfbox{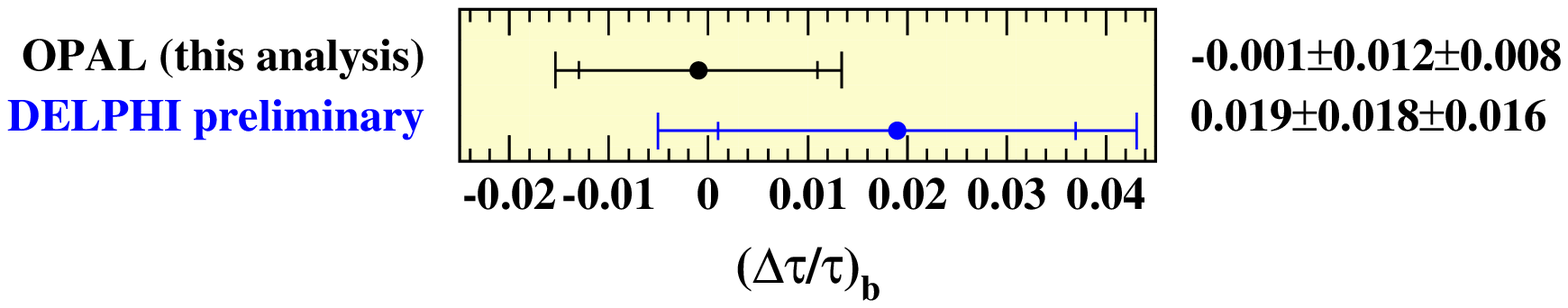}}   
\vskip .2 cm
\caption[]{
\label{fig:worldcp2}
}
\end{figure}

An inclusive sample of ${\rm b}$-hadron decays is used to search for CP
and CPT violation effects. No such effects
are seen. From the time dependent asymmetry of inclusive ${\rm B^0}$
decays, the CP violation parameter is measured to be:
\begin{eqnarray}
Re(\epsilon_B) & = & 0.001 \pm 0.014 \pm 0.003.
\end{eqnarray}
This result agrees with the OPAL measurement using semileptonic b decays:
${\rm Re(\epsilon_B)=0.002\pm 0.007\pm 0.003}$\cite{opaldms}, and is also
in agreement with other less precise results from CLEO and CDF\cite{oldeb}.
The fractional difference in the lifetimes of ${\rm b}$ and 
${\rm \bar b}$-hadrons is also measured to be:
\begin{eqnarray}
\left( \Delta \tau / \tau \right)_b & = &  -0.001
\pm 0.012 \pm 0.008. 
\end{eqnarray}
This is the first analysis accepted for publication which tests the
equality of the ${\rm b}$ and ${\rm \bar b}$-hadron lifetimes.  
These results are summarized in Figures~\ref{fig:worldcp1} 
and~\ref{fig:worldcp2}.

\end{document}